# Guiding Software Developers by Social Networking Application Plug-in using the Multiple Bridge Source Repository through a Data Mining Integrated Approach

Anjela Diana Corraya*, Mousumi Akter Sumi[†], Sadia Islam Shachi[‡] and Ziaur Rahman[§]
Department of Information and Communication Technology,
Mawlana Bhashani Science and Technology University
Bangladesh
Email: *anjela.it12012@gmail.com, [†]sumiakterict@gmail.com, [‡]sadiashachi@gmail.com, [§]zia@iut-dhaka.edu

*Abstract*—In today's world social networking is an important (power full) medium of mass communication. People of almost all classes have been interacting each other and sharing their views, moments and ideas by using enormous user friendly applications in different social networking sites. It's really unbelievable to find a person who never heard about the social network. The available social networking sites usually opportune their users to develop various customized applications through particular templates and embedded sources of codes. The users with average knowledge of development often encounter difficulties to reuse those resources and eventually lack guidelines and necessary API recommendations. In our work, we have proposed a framework and model to help those apps developer through a user assistance plug-in tool that is able to provide identical API usage patterns and sequences in response to a particular user query. We have titled our system as Social Networking Application Plug-in (SNAP). We search social networking apps repository where multiple storage are bridged and apply respective mining algorithm to find the relevant sequences to fulfill the user needs. It provides similar, most relevant and functional API usage scenarios as well as gives an option to choose, reuse and modify the recommended sources. From investigations we have ever made, our SNAP approach is capable to recommend users' error-free, understandable and minimal API patterns.

## I. INTRODUCTION

Nowadays, even a general user wants to build own micro application. Social networking provides opportunities for a user to fulfill his goal. But it is not easy for an inexperienced user to work frequently with different types of class, method, object as well as pattern for different micro application. There are a number of similar applications already existed, which can help the user to easily complete his application. However, existing ways of matching the similarities and taking suggestions from existing repository are quite complicated process. If we use general search engine to find the expected matches, then we will have to deal with huge search results and it is quite difficult to find which one is more relevant. And, in that case, we get recommendation from a repository that does not only contain source codes but also unwanted contents. To solve all of these problems, we have designed a system called SNAP (Social Networking API Plug-in). It presents before the user, the most identical APIs and give the user a choice to choose an eligible type. The great advantage of our plugin is not only filtering a code by its type, but also with its pattern and environment. The operations of plugin are divided into several parts from user query to user recommendation. The most unavoidable parts are filtering and mining. Filtering that means filtering algorithm that finds out flexible pattern matching in strings. Filtering algorithm [1] needs verification from another algorithm for position matching. As we have used mining algorithm after filtering, we choose PrefixSpan [2] mining algorithm for mining. It brings better results comparing with other. This plugin consequently uses this mining algorithm until it brings relevant the code skeleton to the user. So, from the investigating, we have found that the SNAP system brings better result than others.

## II. BACKGROUND AND RELATED WORKS

There are so many works which are related to our works such as: MAPO [3], MAC [4], PR-Miner [5] and so on. MAPO is related to the API mining. As MAPO returns a code segment from the code search engine, the recommended lists may not be so informative. MAPO uses just mining algorithm so it can hardly filter out the same pattern matched code snippets. MAC and MAPO are similar but MAC has the pattern database that is more relevant and suggests the code snippets on work-space. PR-Miner is also a data mining technique; it just extracts implicit programming rules from large software code and detects violations. But it cannot give the best code snippets as different types of coding technologies are merged day by day. Users need to cope with the trade, so extracting of traditional code snippets may not be so much resourceful. Considering the unsolved issues in different existing works, we have tried to design a system to improve the performance. This system uses multiple bridged repositories and also has the offline repository; together are called the SNAP repository. Combination of all these resources make SNAP repository so much resourceful. Filtering and mining algorithms make the SNAP system so much faster and give a better recommendation to the users.

### III. THE PROPOSED APPROACHES OF SNAP SYSTEM

We have developed a novel plugin based on existing code and a frequent sequence miner. In Fig. 1 we want to present our complete overview of our plugin. To get a clear idea about our SNAP system's working flow, we have split whole system into five sections from user authentication to user recommended list. It consists of: user query search, multiple bridged SNAP repository, filtering, frequent sequence miner and re-factoring and updating. Now we illustrate each of the components of our plugin in details:

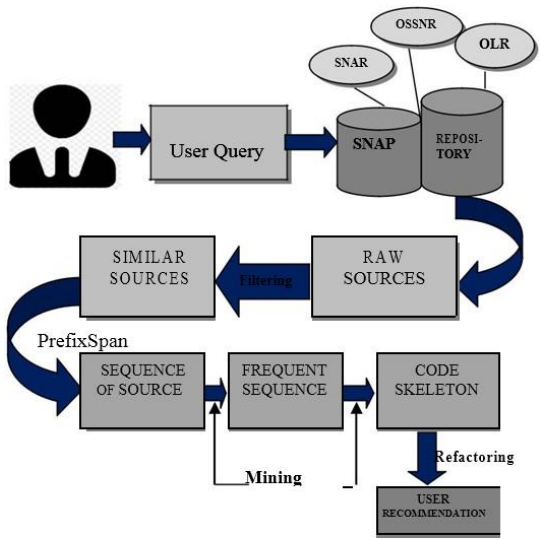

Fig 1. Social Networking APT Plug-in (SNAP) Framework

#### A. User Query Search

At the beginning of the system, a user must have to write a line or codes that will be taken as user query. User writes query on social networking IDE such as Nuclide [6], Facebook Power editor and so on. IDEs must have to open on browser, in there, codes will be typed. The plugin works for following social network: Facebook[7] , MeetMe [8], Orkut [9] , linkdIn [10] and so on. Our user must have internet connectivity or have to connect over LAN and have to use a computer-systematic browser. Now the user have to login to one of those particular social network services. User must be an authenticated and valid as social network user. When a user writes a query on any kinds of social networking editor, then our plugin will give some recommendations, if our plugin is add-ons in that particular users browser. It depends on user whether the user uses our recommended suggestion or not

#### B. SNAP Repository

We have built an idea of our own multiple-bridged repository [11] called SNAP Repository. The plugin starts work as user wish to use our suggestion; we match queries with our SNAP repository according to the query. The searching procedure of SNAP Repository depends on source snippets such as: types of apps, user information and others. Our SNAP repository is a combination of Social Network API Repository (SNAR), Open Source Social Network Repository (OSSNR) and Offline Repository (OLR). SNAR contains the only API for social network where OSSNR depends on Internet search engine and OLR is a state where the computers are connected with each other and create a network and shrinks out the API for social network. But the interesting part is that the codes snippets are searched from offline repository at first. If the recommendations are not acceptable to the users, then the others repositories will be active. So, all these results given by those repositories are called SNAP repository which returns a bunch of relevant codes according to the query for social network.

#### C. Filtering

The number of codes from SNAP repository may be ten thousands or more or less, may be same, similar and dissimilar are found that depends on user query and other actions of our SNAP repository. Our SNAP works for each and every user provided query and returns a collection of codes which called raw sources. Thousands of code snippets can be given by our SNAP repository which is not helpful at all for the users. These raw sources may include a number of same codes used by various famous programmers of one query. Our plugin obtains only the similar sources by filtering it from the raw sources. Filtering algorithm is used for filtering out the portions of codes that are not matched with pre and post codes. So, here arises a thought of position matching. For this, filtering algorithm need co-operation of another position matching algorithm, we recommend here BPM (Bit Pattern Algorithm of Myers)[12]. So, thousand numbers of codes are come in a hundred numbers of codes. The numbers of codes are decreased.

For example the user writes a query one by one under Context class for adding an action to a group in a menu on an editor which is opened on a browser. The SNAP repository shrinks out the all similar codes from other repositories. Now, the filtering algorithm filters all the similar codes and exists only those codes which are not matched with one another. Lets consider three similarities have found for that particular query such as:

- manager.appendToGroup
  (GEFActionConstants.GROUP REST,action),

- manager.appendToGroup
  (GEFActionConstants.GROUP UNDO,action) and
- manager.appendToGroup
  (GEFActionConstants.GROUP VIEW,action)

under Context class which are for custom action, undo action and view action simultaneously.

### D. Frequent Sequence Miner

Hundred numbers of codes are not helpful for developers at all. So, we need to help developers by decreasing these numbers of codes. The next parts of our framework are sequence of source and frequent sequence. These consequently use Sequence Mining Algorithm (PrefixSpan) [2] for mining the codes and arrange codes in accordance with highest uses. Sequence of source is the result by decreasing codes from hundred numbers of codes to fifty code snippets and frequent sequence is the result of decreasing these numbers of codes into ten to five codes for assumption. So, we favor PrefixSpan mining algorithm as it gives attention on frequent prefixes rather than frequent sub-sequences and for each database, the sequential patterns grow by exponential theorem. PrefixSpan mining algorithm scans only prefix sub-sequences; the results must be in form of corresponding post-fix sub-sequences. Then user gets desired code called code skeleton. But the final result from frequent sequence as suppose as Context class from Fig. 2 may not be the desired result for the user which gives the opportunity to view. There need to modify the code. For this, we need another step to help users.

```
public class Context{
    public void Menu(SocialMenuManager manager){
        GEFActionConstants.addStandardActionGroups(manager);

        IAction action;
        action = actionRegistry.getAction(ShowMethodSignatureAction.TEXT);

        if(action.isEnable())

        manager.appendToGroup(GEFActionConstants.Group_VIEW,Action);}}
```

Fig 2. Code snippets

### E. Re-factoring and updating

The target of SNAP plugin is to satisfy the users. The codes from frequent sequence may not help them as there must be needed adding or removing the methods from the class. Suppose, user need to undo an action from the Facebook group, then the contents of Context class that is recommended cannot help the users. Here, users have to add another action. So, the user modifies the class and gets desired output. So, the re-factoring and updating is the last step to get actual code for the users which is more effective than other search engine.

## IV. EVOLUTION

As we are inexperienced, the statistical result of our SNAP system is not professional. We take some common environmental setup, common knowledge and use two different ways to get the effectiveness of the system.

### A. Environmental Setup

In order to show the evolution of our plugin approach, we are used following environmental tools.

- 5 web based API using projects,
- Particular search engine belongs the social network (Koders.com [13]),
- Bridge Repository (SNAP repository)

### B. Effective Evolution

In order to evaluate the effectiveness of our plugin by comparing with other existing search engine like Koders.com. We have compared in two ways. They are Effectuate Evaluation as well as Empirical Study.

For Example, on the following Table 1, we are taken 5 numbers of users Query. We use those on both Koders.com and on SNAP. And then we listed the number of matches of that particular query on the chart. Now on Fig. 3 we are going

Table I
EFFECTIVENESS EVALUATION

| Serial | User Query | Koders.com | SNAP |
|---|---|---|---|
| 01 | GEFActionConstants.GROUP UNDO,action | 100 | 30 |
| 02 | GEFActionConstants.GROUP VIEW,action | 120 | 20 |
| 03 | HTTP Response.setContent | 80 | 10 |
| 04 | HTTP Web Request.getResponse | 90 | 30 |
| 05 | SQL Connection open | 120 | 80 |

to analysis these comparisons using graph.
From the analysis of this following graph Fig. 3 we can say

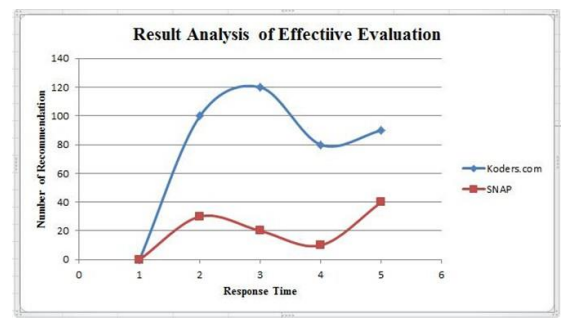

Fig 3. Result Analysis of Effective Evolution

that SNAP is much more effective than Koders.com.

*C. Empirical Study*

On the empirical way, we are taking two groups, Team A and B where Team A's searching process based on Koders.com and Team B uses SNAP. Then we give each of the group several numbers of different API queries. In Tables 2, we listed the number of facing problems of each of the teams to see the better comparison result.

Now on Fig. 4 we are going to analysis these comparison results using graph.

Table II
EMPIRICAL STUDY

| Serial | Number of User Query | Team A(Koders.com) | Team B(SNAP) |
|---|---|---|---|
| 01 | 15 | 10 | 5 |
| 02 | 12 | 6 | 3 |
| 03 | 14 | 7 | 4 |
| 04 | 17 | 5 | 3 |
| 05 | 19 | 6 | 5 |

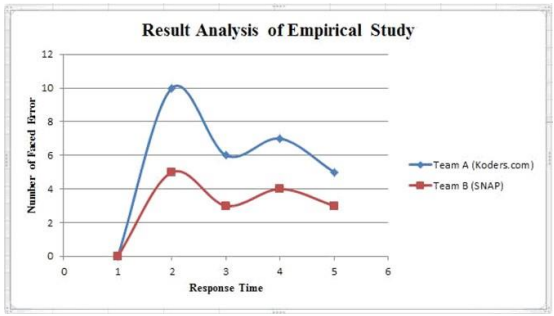

Fig 4. Result Analysis of Empirical Study

From the above graph we can say that SNAP system faces less difficulty and error then Koders.com. From the following two figures we can get the clear idea about the effectiveness of SNAP Plugin.

## V. LIMITATIONS AND THREATS TO THE EVOLUTION

SNAP plugin is not self-reliance. It has boundaries to help the developers. It works only on social networking editor and can help only micro-apps developers. This plugin is able to extract API information which may not be enabled to compile and debug. It is unable to give suggestions on more than one pattern at a time. It works on first pattern which is inputted by users. In case of second pattern the process will be restarted. We also have some limitations on behalf of the empirical study. Like,

- Analytically the result will be varied by the no. of user queries.
- The Team members also have affects on the result.
- Mining Strategy is also a big concern for the whole process. Its correction is very much endeavor for the effective result.

## VI. CONCLUSION AND FUTURE SCOPES

SNAP plugin helps the normal users to develop their own micro-apps. When the user searches for codes on different kinds of search engines [14], they find lots of wanted and unwanted codes which may not help the user to get actual helpful codes. SNAP plugin helps them to carry out those useful codes and help them to develop their own apps. The overall investigation of our paper is for helping the Social networking apps developer in the exact way according to their demand. SNAP plugin is not only mining API-usage patterns with accuracy but also filtering as well as matching with the types belongs to similarities. In order to help developer, the plugin is locating significant and agreeable code snippets in the particular query with more sincerity. The ascertained of SNAP is that it provides code as a recommended list with lower bugs. In future, we will design our own repository which contains a huge collection of queries. Next, we will extremely focus on our empirical study to build it up without limitations. We need dynamic investigation to make sure that our selected technique is the best or not. In future we will try to upgrade our system on the basis of professional implementation. We also try to improve the whole evaluation process. Our main future preferable work is to build a standard repository in our SNAP system.